\documentclass[sigconf]{acmart}
\usepackage{amsmath,bm}
\usepackage{xspace}

\usepackage{mathtools,amssymb,lipsum}
\usepackage{cuted}
\usepackage{caption}
\usepackage{subfig}
\usepackage{kantlipsum,afterpage,hyperref} 
\usepackage{listings}
\makeatletter
\def\thm@space@setup{\thm@preskip=0pt
\thm@postskip=0pt}
\makeatother
\newtheoremstyle{newstyle}      
{} 
{} 
{\mdseries} 
{} 
{\bfseries} 
{.} 
{ } 
{} 

\theoremstyle{newstyle}
\newtheorem{thm}{Theorem}[section]


\AtBeginDocument{%
  \providecommand\BibTeX{{%
    \normalfont B\kern-0.5em{\scshape i\kern-0.25em b}\kern-0.8em\TeX}}}
    
\newcommand{\mname}{\texttt{Split-Treatment}\xspace}
\definecolor{emk}{RGB}{40, 134, 153}

\newcommand{\indep}{\perp \! \! \! \perp}

\setcopyright{rightsretained}
\copyrightyear{2018}
\acmYear{2018}
\acmDOI{10.1145/1122445.1122456}

\acmConference[Woodstock '18]{Woodstock '18: ACM Symposium on Neural
  Gaze Detection}{June 03--05, 2018}{Woodstock, NY}
\acmBooktitle{Woodstock '18: ACM Symposium on Neural Gaze Detection,
  June 03--05, 2018, Woodstock, NY}
\acmPrice{15.00}
\acmISBN{978-1-4503-9999-9/18/06}

\begin{document}

    \title{\mname Analysis to Rank Heterogeneous Causal Effects for Prospective Interventions} 


\author{Yanbo Xu}
\affiliation{%
  \institution{Georgia Institute of Technology}
}
\email{yxu465@gatech.edu}

\author{Divyat Mahajan}
\affiliation{%
  \institution{Microsoft Research India}
}
\email{t-dimaha@microsoft.com}

\author{Liz Manrao}
\affiliation{%
  \institution{Microsoft}
}
\email{elmanrao@microsoft.com}

\author{Amit Sharma}
\affiliation{%
  \institution{Microsoft  Research India}
}
\email{amshar@microsoft.com}

\author{Emre K\i c\i man}
\affiliation{%
  \institution{Microsoft Research AI}
}
\email{emrek@microsoft.com}








\begin{abstract}
For many kinds of interventions, such as a new advertisement, marketing intervention, or feature recommendation, it is important to target a specific subset of people for maximizing its benefits at minimum cost or potential harm. However, a key challenge is that no data is available about the effect of such a {\em prospective intervention} since it has not been deployed yet.
%
In this work, we propose a {\em split-treatment} analysis that ranks the individuals most likely to be positively affected by a prospective intervention using past observational data.
Unlike standard causal inference methods, the split-treatment method does  not need any observations of the target treatments themselves.  
Instead it relies on observations of a  {\em proxy treatment} that is caused by the target treatment. Under reasonable assumptions, we show that the ranking of heterogeneous causal effect based on the proxy treatment is the same as the ranking based on the target treatment's effect.
%
%
%
In the absence of any interventional data for cross-validation, \mname uses sensitivity analyses for unobserved confounding to select model parameters.
We apply \mname to both a simulated data and a large-scale, real-world targeting task and validate our discovered rankings via a randomized experiment for the latter. 


\end{abstract}

\maketitle

\section{Introduction}


Identifying individuals who will benefit from an intervention or treatment is an important challenge in many domains.
We focus on this problem in the context of computing applications.  For instance, many modern applications and devices display educational messages or other recommendations to users, telling them about undiscovered features, advertising related products, or providing other tips that may aid them.  
In complex software and scenarios, not all such messaging interventions are useful for every individual, and some may even be detrimental.  For example, a message recommending an advanced feature may provide significant help to some advanced users, but also confuse and thus harm others.  

There are several practical approaches to identifying individuals who will benefit from a specific messaging intervention based on reinforcement learning, such as contextual bandits~\cite{li2010contextual} and off-policy learning~\cite{gilotte2018offline,swaminathan2017off}. However, these approaches face limitations when applied to targeting \textit{prospective} interventions where the potential harms or other costs of the intervention limit our ability to run experiments. Contextual bandits require active experimentation for every new message or intervention to select the individuals to assign treatment. That is,  they may require exposing individual users to detrimental messages and causing negative effects at scale, including negative effects of irrelevant ads like ad blindness~\cite{resnick2014impact} and annoyance; and negative effects of users taking action on mismatched recommendations like user confusion and frustration. In addition, they depend on  feedback (reward) signals in relatively shorter time-frames, such as click signals for online ads. However in many setting, the outcome of interest is longer-term such as sustained usage of a product or a feature. While off-policy learning does not require experimentation, it assumes that the target treatment has been deployed and therefore observed data for the intervention is available. Therefore, existing approaches either require active experimentation
or assume that a new intervention's effects can be estimated based on the effects of past interventions. 

%

\begin{figure}
    \centering
    \includegraphics[width= .3\textwidth]{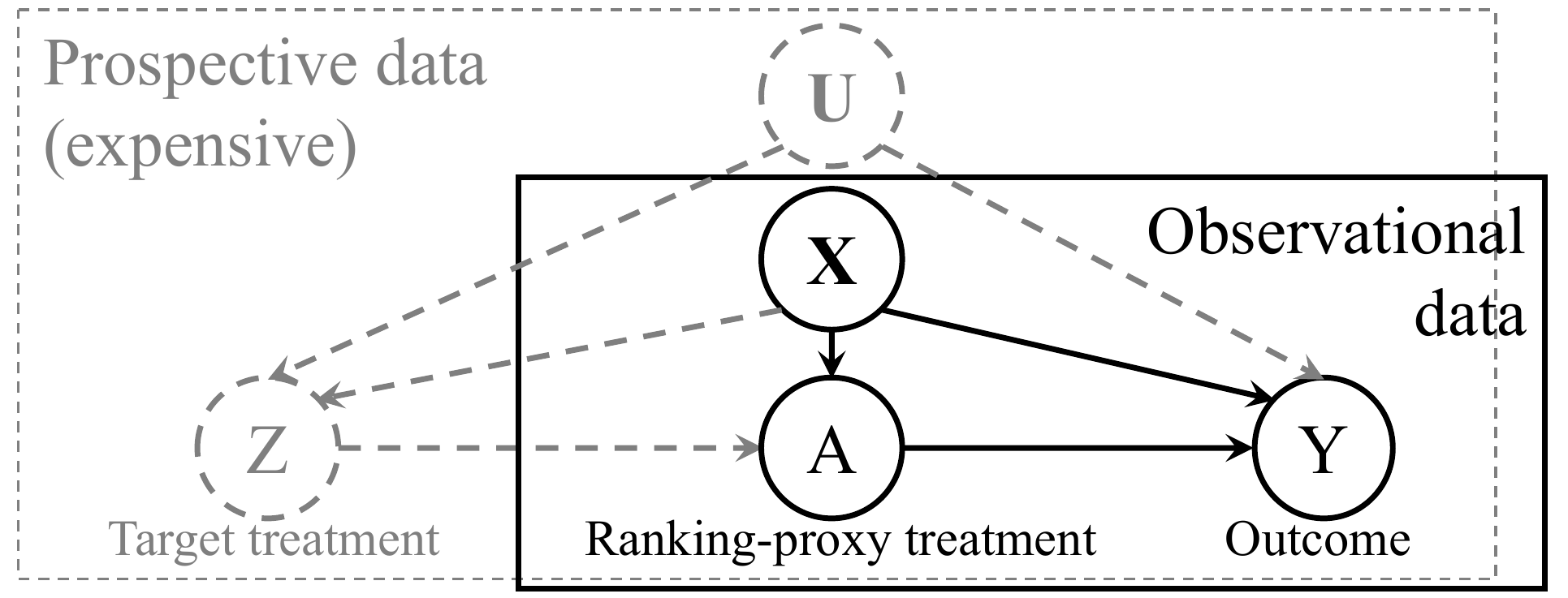}
    \vspace{-.5em}
    \caption{The goal is to estimate the effect of a prospective treatment that is never observed. \mname uses a proxy treatment $A$ that lies on the causal path between $Z$ and $Y$. Observed variables ($\bm{X}$) confound the effect of $A$ on $Y$ (and possibly also $Z$'s effect). Under proper assumptions, individuals for which the proxy treatment has a positive effect will also have a positive effect to the target treatment.   
    }
    \label{fig:split_treatment}
    \vspace{-1em}
\end{figure}
We present and evaluate an alternative approach that uses observational (non-experimental) data for identifying individuals who are likely to benefit from a prospective, novel intervention (Fig.~\ref{fig:split_treatment}). 
To do so, we realize that it is often possible to split the effect of a messaging treatment into two parts: (1)~how persuasive an treatment is in encouraging an individual to take a short-term action; and (2)~the long-term effect of the action.
While we cannot know the first part of our split-effect without performing the prospective intervention, we can identify the second part of our split-effect when the action we are encouraging people to take has been performed by others in our non-experimental data.  
We show that, in many situations, this is sufficient for ranking individuals most likely to benefit from the otherwise novel intervention. %
Beyond this analytical approach, we make the additional contribution 
of demonstrating how challenges that commonly threaten the validity of observational studies can be addressed in this setting through the careful application of refutation and sensitivity analyses.

We apply \mname to estimate the heterogeneous causal effect of a novel messaging intervention in a large application suite. 
Each messaging intervention, $Z$, is a recommendation for a feature, $A$.  The outcome of interest is the feature's long-term adoption, $Y$.
Our method enables ranking of individuals based on the estimated effect of the intervention. We implement methods for causal effect estimation based on inverse propensity weighting and machine learning models like random forests and CNN, and use sensitivity analysis to arrive at robust models. Finally, we validate these models based on a randomized experiment and demonstrate that our methods and sensitivity analyses correctly identify, {\em a priori} the most accurate of the ranking models.
We make 3 contributions: 

\begin{enumerate}
    \item {\bf Split-Treatment:}  We propose an identification technique for \textit{ranking} the effect of a prospective treatment without any access to the data with the treatment. We contrast our approach with instrumental variable and front-door analyses that require observed data for the treated population. 
    
    \item {\bf Sensitivity Analysis:} How can we practically check the assumptions and ensure reliable conclusions from our observational data?  We show how to adapt sensitivity methods for this problem and demonstrate that their violations correlate with empirical errors.  
    
    \item {\bf Validation through Active Experiments :} We validate our conclusions using an A/B experiment  of recommendation efficacy for a feature on over 1M users. 
    Our findings show that \mname rankings of the individuals that are most likely to benefit from the recommended message matches the results from A/B experiments. 

\end{enumerate}

\section{Background and Related Work}
\subsection{Causal analysis}




Conventionally, the problem of showing recommended items to people is considered an outcome prediction problem: what would be the predicted outcome (e.g., rating) for an item by an individual? Based on this prediction, a ranked list of recommended items is shown to the individual~\cite{su2009survey}. However, recent work~\cite{sharma2015estimating, schnabel2016recommendations} frames recommendation as a causal inference problem: what is the effect on the outcome (e.g., usage metrics) of showing a recommended item to an individual?  Rather than simply verifying if a person achieves the predicted outcome metric, it is important to understand if the desired outcome was achieved \emph{because} of the recommendation. If the outcome would have been the same without the recommendation, then it does not have a causal effect on the outcome. 

 One of the fundamental challenges in causal analysis is we only observe one of the outcomes: either an intervention was shown, or not. 
Following the Structural Causal Model (SCM) described in \cite{pearl1995causal,judea2000causality}, we can define the interventions and counterfactuals through a mathematical operator called $do(z)$: $Y|do(z=1)$ refers to the outcome measure had intervention (or treatment) Z enforced, and $Y|do(z=0)$ refers to the  outcome had no treatment. The causal effect of treatment Z is:
$$\texttt{Causal Effect} = \mathbb{E}[Y|do(z=1)] - \mathbb{E}[Y|do(z=0)].$$
\noindent To address the causal question, one needs to account for \emph{confounding} in the observed data. 
Formally, confounding is due to the intervention (recommendation) and outcome having common causes. In general, we expect a treatment to have varying effects across individuals. There is recent research~\cite{wager2018estimation,chernozhukov2017double,oprescu2019orthogonal,kunzel2019metalearners} on how to estimate heterogeneous causal effect for different subgroups, known as the Conditional Average Treatment Effect (CATE). In the limit, we may consider the causal effect for each individual, known as the Individual Treatment Effect (ITE)~\cite{shalit2017estimating}. We extend this line of work by proposing a method to rank causal effect for prospective treatments, for which we have no data as yet. 

That said, since counterfactual outcomes are never observed simultaneously, evaluating a causal estimate is non-trivial. Unlike prediction, it is not possible to validate a causal effect estimate based only on observed outcomes. Typically, any causal estimation method makes certain assumptions on the counterfactuals and their robustness depends on how plausible those assumptions are. There is past work on validating assumptions of causal models to the extent possible, often by exploring the sensitivity of the obtained estimate to unobserved variables~\cite{robins2000sensitivity,rosenbaum2010design,diaz2013sensitivity}. We utilize sensitivity analyses to rank different candidate methods for causal estimation.

\subsection{Effects of Ads or Recommendations} 
Previous work for identifying individuals who will benefit from messages, ads, or recommendations mainly fall into two groups. The first group of work \cite{sundar1998does,winer2001framework,resnick2006value,lewis2014online} rely on campaign studies to analyze the effect of the known treatment ads through A/B testing or reinforcement learning-based methods like contextual bandits~\cite{li2010contextual,tang2013automatic,lattimore2016causalbandits} that target treatment to specific individuals during an experiment.
\citet{wang2015robust} employs a tree structure in modeling treatment propensities in observational studies, and also enables analysis of both uni- and multi-dimensional ad treatments. They validate the results through simulation study, and also apply their method in two real campaigns.   
\citet{brodersen2015inferring} applies Bayesian structural time-series models to infer causal impact of market interventions on an outcome metric over time. They illustrate the statistical properties of posterior inference on simulated data, and demonstrate their approach practically in an online ad campaign. By referring to our task as a \textit{unit selection problem}, recent work ~\cite{ijcai2019li} integrates Pearl's structural causal model to formulate a different objective function in evaluations of A/B testing that accounts for counterfactual nature of the desired behaviors (e.g., compilers, always-taker, never-taker and defiers). 
Our work differs from these previous analysis of the causal impact of ads campaigns in that we are trying to predict the impact of an advertisement {\em before} starting a campaign. 



The second group of work do not require treatment ads to be randomized, but need them exposed in the data, and then adjust selection bias in such observational data. For example, \citet{chan2010evaluating} utilizes doubly robust methods to evaluate the effectiveness of online ads from large observational data, and validates results using simulations based on realistic scenarios.
\citet{gordon2019comparison} assesses empirically how the variation in data availability can affect the ability of observational methods to recover the causal effects of online advertising learnt from randomized trials. 
Our work differs from these observational studies in that we estimate the impact of a prospective message that does not exist in the current data.




\section{Identification by \mname}
\label{sec:formulation}

In this section we describe how the \mname method can identify the ranking of individuals most likely to benefit from a novel treatment or recommendation. 
Suppose $Z$ is the prospective treatment of interest and $Y$ is the outcome we aim to measure its effect on. If we run an A/B experiment where half of the population is treated with $Z$, then we can assess the conditional average treatment effect (CATE) of $Z$ within a subpopulation $\mathcal{G}$ by taking the difference between the two outcome means respectively measured in the treatment and control groups:
\begin{equation}
\text{CATE}^{(z)}_{\mathcal{G}}= \mathbb{E}_{\mathcal{G}|_{Z=1}} [Y \mid Z = 1] -  \mathbb{E}_{\mathcal{G}|_{Z=0}} [Y \mid Z = 0],
\label{eq:ate}
\end{equation}

However, our observational data does not contain the prospective (target) treatment $Z$, as it has not been deployed yet.  To achieve the goal, we propose  splitting the target treatment's effect into two parts.  The first part is the effect of the target treatment on some short-term user behavior $A$, such as the first immediate action encouraged by the treatment $Z$.  The second part is the effect of this immediate action $A$ on the outcome measure $Y$ that we are optimizing. The requirements for our choice of $A$ are: (1) The effect of $Z$ on $Y$ should be mediated through $A$.  That is, $Z$ does not directly affect the outcome;  (2) $Z$ does (or would) effect the use of $A$; (3) $A$ exists, with some natural variation, in our observational logs. 

Note that the first two requirements are exactly the requirements for a valid instrumental variable (exclusion and relevance~\cite{angrist1996identification}), except in this case $Z$ is not observed. In addition, $Z$ need not be a valid instrumental variable. It may be caused by the same variables that confound the effect of $A$ on $Y$ (however there cannot be any unobserved common cause for $Z$ and $A$). For example, the prospective intervention may be assigned based on the features of each individual. The causal graph is also similar to the front-door identification criterion (Definition 3.3.3, ~\citet{judea2000causality}), however the criterion cannot be directly applied since $Z$ is unobserved. 
Instead we use the following assumptions to use $A$ as a proxy for ranking the most promising individuals for assigning the target treatment.

\noindent \textbf{Assumption 1} (Ignorability): There exists no unobserved confounding between the proxy treatment $A$ and the outcome measure $Y$:  $P(Y|do(a), \bm{x})= P(Y|a, \bm{x})$, 
 for some observed variables $\bm{X}$.



\begin{thm}
 Under the causal graph $G$ from Figure~\ref{fig:split_treatment} and given Assumption 1, the post-intervention distribution of $Y$ resulting from the intervention $do(z)$ conditional on $\bm{x}$ can be identified as, 
$$P(Y|do(z), \bm{x}) = \sum_a P(a|z, \bm{x} )P(Y|a, \bm{x}).$$
\end{thm}
\noindent \textit{Proof.}
For any set of nodes $Y$, $R$, $S$ and $T$, the rules of the do-calculus~\cite{pearl1995causal} can be written as (where $G_{\text{ }\overline{B}/\underline{B}}$ denotes the graph by deleting from $G$ all arrows pointing to/emerging from the set of nodes $B$ respecively).

\noindent \textbf{\textit{Rule 2:}} $P(Y|do(r), do(s), t) = P(Y|do(r), s, t) \text{, if } (Y \indep S|R, T)_{G_{\text{ }\overline{R},\underline{S}}}$;

\noindent \textbf{\textit{Rule 3:}} $P(Y|do(r), do(s), t) = P(Y|do(r), t) \text{, if } (Y \indep S|R, T)_{G_{\text{ }\overline{R},\overline{S(T)}}}$, 
 where $S(T)$ are the $S$-nodes that are not ancestors of $T$ in $G_{\text{ }\overline{R}}$. 
\begin{equation*}
\begin{split}
    P(Y|do(z), \bm{x}) &= \sum_a P(a|do(z), \bm{x}) P(Y|do(z),a, \bm{x})  \\
    &= \sum_a  P(a|z, \bm{x}) P(Y|do(z), do(a), \bm{x})  \quad \text{ ... Using } \textit{Rule 2} \\
    &= \sum_a P(a|z, \bm{x}) P(Y|do(a),\bm{x}) \quad  \quad  \quad  \text{ ... Using } \textit{Rule 3}\\
    &= \sum_a P(a|z, \bm{x}) P(Y|a,\bm{x})  \quad  \quad  \text{ ... Using } \textit{Assumption 1}
    \end{split}
\end{equation*}
Using the above theorem, for individuals characterized by $\bm{x}$, we can write their individualized treatment effect (ITE) of $Z$ as,
\begin{align}
\label{eq:split}
    \text{ITE}^{(z)}(\bm{x}) & = \mathbb{E}[Y|do(z=1),\bm{x}] - \mathbb{E}[Y|do(z=0),\bm{x}]\\\nonumber
    & = \big(\underbrace{P(a =1|z=1,\bm{x})- P(a=1|z=0, \bm{x})}_{
    \text{Compliance($\bm{x}$})} \big) \\\nonumber
    & \quad\quad \cdot \big(\underbrace{(\mathbb{E}[Y|a=1, \bm{x}]- \mathbb{E}[Y | a=0, \bm{x}]}_{\text{ITE}^{(a)}(\bm{x})}\big).
\end{align}
\noindent While we cannot identify the first term,  $\text{Compliance}(x)$, as $Z$ does not exist in our data, we can still estimate $\text{ITE}^{(a)}$ from the observational data.
If we take the group $\mathcal{G}$ containing the most effective individuals to the proxy treatment $A$ to be the same group of the most promising individuals to the target treatment $Z$, we can derive another format of Eq. (\ref{eq:ate}) based on Eq. (\ref{eq:split}) into the following:
\begin{align}
\label{eq:cate}
   \text{CATE}^{(z)}_{\mathcal{G}} & = \mathbb{E}_{\bm{x} \in \mathcal{G}} \big[ \text{ITE}^{(z)}(\bm{x}) \big] \\\nonumber 
   & \propto \mathbb{E}_{\bm{x} \in \mathcal{G}} \big[ \text{ITE}^{(a)}(\bm{x}) \big], \text{ for } \mathbb{E}_{\bm{x} \in \mathcal{G}}\big[\text{Compliance}(\bm{x})\big] > 0.
\end{align}

\noindent \textbf{Assumption 2} (Compliance): We assume the target treatment $Z$ have a positive average effect to encourage the proxy treatment $A$ within the selected group containing the most effective individuals to $A$. That is,  $ \mathbb{E}_{\bm{x} \in \mathcal{G}}[P(a=1|z=1, \bm{x}) - P(a=1|z=0, \bm{x})] > 0$.

We cannot test this assumption in observational data due to the absence of $Z$, so we must use external domain knowledge to ensure that the assumption is satisfied. 

\section{Estimation using \mname}
The key challenge in feature recommendation, such as marketing software features to existing users, is to identify who would benefit from using a feature.  
Here, we show how \mname can be used to rank individuals using only observational data.

\begin{figure}[t]
    \centering
\includegraphics[width=\columnwidth]{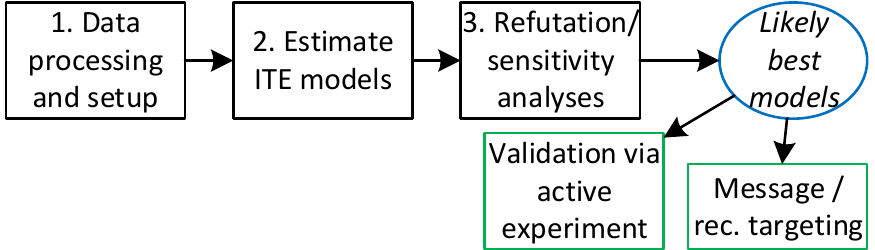}
\vspace{-1em}
\caption{An end-to-end analysis pipeline of using \mname in feature recommendation. Validation is added only if experimental data having $Z$ is available.} 
\label{fig:pipeline}
\end{figure}

%
Following the formulation of \mname in Eq. (\ref{eq:cate}), we say the target treatment $Z$ is the given message recommending a software feature.  We pick the proxy treatment $A$ to be an individual's first trial of the feature; and we define the outcome measure $Y$ 
to be an observation that the individual is receiving benefit from $A$.  In our analysis, $Y$ is sustained usage of the recommended feature measured in a post-treatment window, indicating the individual found the feature useful enough to keep engaging with it.

We split the treatment effect, use only observational data to learn the ITE function $\text{ITE}^{(a)}$ for treatment $A$, and rank the individuals characterized by $\bm{x}$'s as the most likely to keep using the feature if they have tried it now. We select the top K individuals into the recommendation group $\mathcal{G}$. By making the assumption that a well-designed message will encourage these individuals to try out the product (i.e., positive compliance of $A$ to $Z$ in the selected group), we claim that the individuals in group $\mathcal{G}$ are most likely to be positively affected by the recommendation such that they will use the feature and sustain using it once they got exposed.   


Our end-to-end observational analysis and validation pipeline is shown in Figure~\ref{fig:pipeline}. 
First, we collect and analyze observational data containing the proxy treatment $A$; we use this to estimate the function $\text{ITE}^{(a)}$.  Using this ITE function, we rank individuals by the effect that $A$ (and thus in turn, $Z$) will have on them.   
In our analysis of this data, we might use different algorithms, and even different feature sets.  
%
Secondly, we test our causal assumptions through sensitivity analyses.  These help us identify the particular analysis design that appears to be most robust.
Finally, in our validation stage, we use experimental data to  
check that our \mname observational analysis 
has correctly identified the recommendation group $\mathcal{G}$.  Note that in practice, experimental data for validation will not be available for all prospective treatments. As we will show in the empirical analysis,  sensitivity analyses can be useful to select robust models in the absence of experimental validation (or when experiments are available only for a limited number of interventions). 
Below we provide further details of the three major components in the pipeline.  

\subsection*{Step 1. Data preprocessing and set up}

Given a cohort of users,
we divide the observational data temporally into three windows.  
Observations made during a \textit{pre-treatment window} provide a behavioral baseline of users ($X$).  We use this $X$ to address confounding.
We look for observations of $A$ during the \textit{treatment window}, labeling a user as treated if we observe $A$, and untreated otherwise.  
%
In the \textit{post-treatment window}, we measure the outcome $Y$ that indicates a user has benefited from the action $A$.
%

\begin{figure}[t]
    \centering
    \includegraphics[width= .45\textwidth]{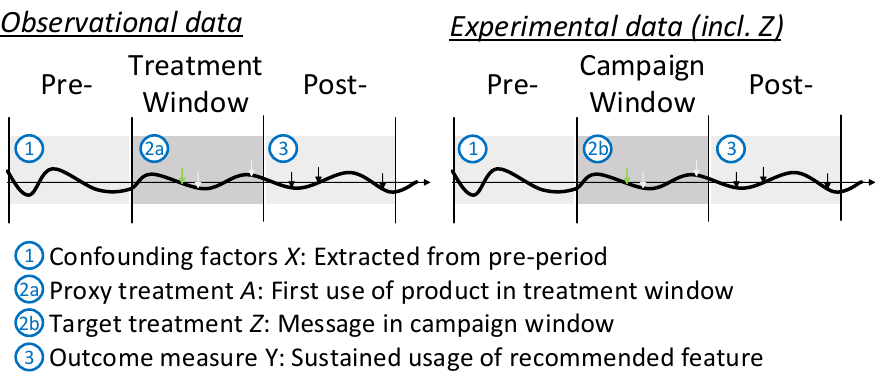}
    \caption{Data timeline in observational and experimental regimes
    }
    \label{fig:data-timeline}
\end{figure}

For target data that our proposed method will be deployed on, we will only need to extract the user features $X$ in the Pre-treatment window prior to the time point when the feature recommendation decision should be made. 
%
%
But for experimental data that we use for validation, we need to align the data timeline between the observational and experimental regimes. Instead of specifying a Treatment window, the experimental data has a \textit{Campaign window} during which the ad treatment $Z$ was randomly assigned and actively tested; the proxy treatment  $A$ of the first usage is also measured in this window. We exclude those users who still got exposed to the ad in the \textit{Post-campaign window}, and count the sustained usage $Y$ in that window. The user features are again extracted from the \textit{Pre-Campaign window}. The three windows are aligned with the same length between the two regimes, as shown in Figure~\ref{fig:data-timeline}.

\subsection*{Step 2. Estimating Individual Treatment Effects}




The key challenge of estimating causal effects from observational data is to adjust the factors $X$ that can potentially confound the causal relationship between treatment $A$ and outcome measure $Y$. 
Given unconfoundedness in Assumption 1, existing causal methods fall into three main categories: inverse probability of treatment weighting (IPTW) based methods that predicts propensity scores ($P(a=1|\bm{x})$) to reweight individuals and obtain unbiased ATE estimates; g-formula based methods that predict the two arm potential outcomes $\mathbb{E}[Y^{(a)}|\bm{x}]$ for individuals and obtain unbiased ITE estimates; doubly robust methods that combine the two methods so that only one of the model need to be correctly specified. We develop a generic doubly robust framework in \mname by fitting a logistic regression model for generating propensity scores and using different machine learning techniques for estimating the potential outcomes. Our regression choices include Poisson Regression, linear regression with Stochastic Gradient Descent, Fast Tree Regression, Fast Forest Regression, and a 2-layer Convolutional Neural Network (CNN).  These methods can be switched to classification models if the outcome measure is binary. 

We define an outcome regression function $f: \mathbb{R}^K \times \{0,1\} \rightarrow \mathbb{R}$, where $K$ is the cardinality of the input feature $X$ and treatment $A$ is considered binary, and compute ITE for an individual $\bm{x} \in \mathbb{R}^K$ as 
\begin{align}\text{ITE}^{(a)}(\bm{x}) \equiv f(\bm{x}, a = 1) - f(\bm{x}, a = 0).
\end{align}
Given observational data $\mathcal{D}^{(n)} = \{(y_i, \bm{x}_i, a_i)\mid_{i=1}^n\}$, we can learn $f$ by minimizing the following loss function
\begin{align}
\sum_{i=1}^n w_i \mathcal{L} (y_i, f(x_i, a_i)),
\label{eq:w}
\end{align}

\noindent where $\mathcal{L}$ is a Poisson loss in Poisson regression and CNN, and mean squared error for the other regressions; $w_i$ is the IPT weight computed from the predicted propensity score.  Note that loss in Eq. (\ref{eq:w}) goes back to ordinary regression loss if $w_i \equiv 1$, and becomes no longer a causal model with confounding adjustments. In particular, we use the Stabilized IPTW \cite{robins2000marginal} to compute the weight $w_i$: 
\begin{align}
w_i = a_i \cdot \frac{ P(a_i = 1)}{P(a_i = 1|\bm{x}_i)} + (1-a_i) \cdot \frac{1- P(a_i = 1)}{1-P(a_i = 1|\bm{x}_i)}.
\label{eq:sw}
\end{align}
\subsection*{Step 3. Sensitivity Analyses}
\label{sec:refutation}
In general, multiple treatment effect estimation methods will provide varying results about the causal effect. These can be due to faults in different parts of the analysis pipeline: mis-specification of the underlying causal model, due to errors in identification, or due to limits in estimation. Here we provide methods to evaluate the estimates returned by different methods. Note that similar to statistical hypothesis tests, the tests below can refute an estimator that may be unsuitable, but cannot find the ``correct'' estimator. In other words, we can weed out some unreliable estimators, but cannot prove that a certain estimator is the closest to the true estimate.

We consider two tests: \textit{placebo} test and \textit{unobserved-confounding} test. In the first, we introduce a random variable and rerun the analysis assuming it is the treatment variable~\cite{athey2017state,dowhy}. Our expectation is that an estimator should return a zero causal effect. To the extent that estimate varies significantly from zero, we can assess the bias of the estimator and prune out estimators that show substantial bias. In the unobserved-confounding test, we wish to estimate how sensitive the models are to the presence of unobserved confounders. In practice, it is plausible that some confounders were missed, so we would like to prefer methods whose estimates are relatively stable in the presence of new confounds, especially not vary substantially with small changes in confounding.   To verify this,  we add a new confounder to the feature set with varying degrees of its effect on $A$ and $Y$, and re-compute the causal effect. While we expect the causal estimate to change as the degree of effect of the confounder is increased, a better estimator is expected to be less sensitive to such changes. Thus, we can rank estimators based on the variation in their estimate for the same amount of confounder's effect added. 

To implement the second test, we model the confounder as sampled from a Gaussian distribution and use the Bayes rule to arrive at the posterior distribution that shows correlation with both the outcome and the treatment. 
Let us denote the new confounder as $U$, treatment variable as $A$ and the outcome as $Y$. Intuitively, we choose the degree of effect of $U$ on $Y$ and $A$ to a desired value, and then use Bayes Rule to obtain the distribution of such a $U$.  
\begin{align*}
P(U|Y,A)=\frac{P(Y|U,A)p(U|A)}{P(Y|A)}
\end{align*}
Since we are interested in modeling the direct effect of $U$ on $Y$, we can ignore the causal association between $Y$ and $A$ and obtain, 
\begin{align*}
P(U|Y,A)=\frac{P(Y|U)p(U|A)}{P(Y)}
\end{align*}

For our experiments, we use a parametric form for the distributions. To implement the relationship between U and A, $P(U|A)$,  we assume the following the gaussian prior on $U$: $U \sim \mathcal{N}(u | u_{0},\epsilon$); where $u_{0}= \alpha + a$, depending on a given realisation of treatment variable $A=a$ and tunable parameters $\alpha$ and $\epsilon$. For the relationship between Y and U, we then define the likelihood of observing all the outcome samples  using gaussian distribution as follows: $\prod_{i=1}^{N_{a}}\mathcal{N}(y_{i}|u, \epsilon)$ where $N_{a}$ are the total of data points with treatment $A=a$, and $\epsilon$ is a tunable parameter denoting the strength of effect (higher $\epsilon$ corresponds to a weaker effect of $U$ on $Y$) . 
Substituting the gaussian prior and the likelihood, we obtain:
\begin{align}
    \mathcal{N}(u | u^{*}, \epsilon^{*})=     \frac{(\prod_{i=1}^{N_{a}}\mathcal{N}(y_{i}|u, \epsilon))*\mathcal{N}(u | u_{0},\epsilon)}{ \int_{}^{}  \prod_{i=1}^{N_{a}}\mathcal{N}(y_{i}|u, \epsilon) du}
\label{eq:ref-analysis}
\end{align}

\noindent Based on the above solution, we obtain,

$$u^{*}= \frac{u_{0}+N_{a}*\sum_{i=1}^{N_{a}} y_{i}}{N_{a}+1} \quad \epsilon^{*}= \frac{\epsilon}{N_{a}+1}$$

\noindent An important to notice is that the posterior distributions $\mathcal{N}(u | u^{*},$ $\epsilon^{*})$ is dependent on the treatment value via $u_{0}= \alpha + a$. Hence, for each data point; we sample value for the new confounder from the posterior distribution corresponding to the particular treatment class. The mean of the posterior distribution is a weighted combination of the prior mean ( dependent on treatment ) and the average of the outcome variables; which ensure the new confounder $U$ is correlated with both $A$ and $Y$. 

Once the confounder is generated, we re-compute the estimate with this additional confounder and compare the estimate to its original value for each method. Since our goal is to select individuals for treatment (rather than estimating individual treatment effect), we evaluate each method on what fraction of the individuals selected for treatment stay the same between the original estimate and the new estimate. Specifically, we consider individuals whose estimate was higher than the median in both analyses, and compute the number of common individuals between the two analyses.  




\begin{figure}[tb]
    \centering
    \includegraphics[width= \columnwidth]{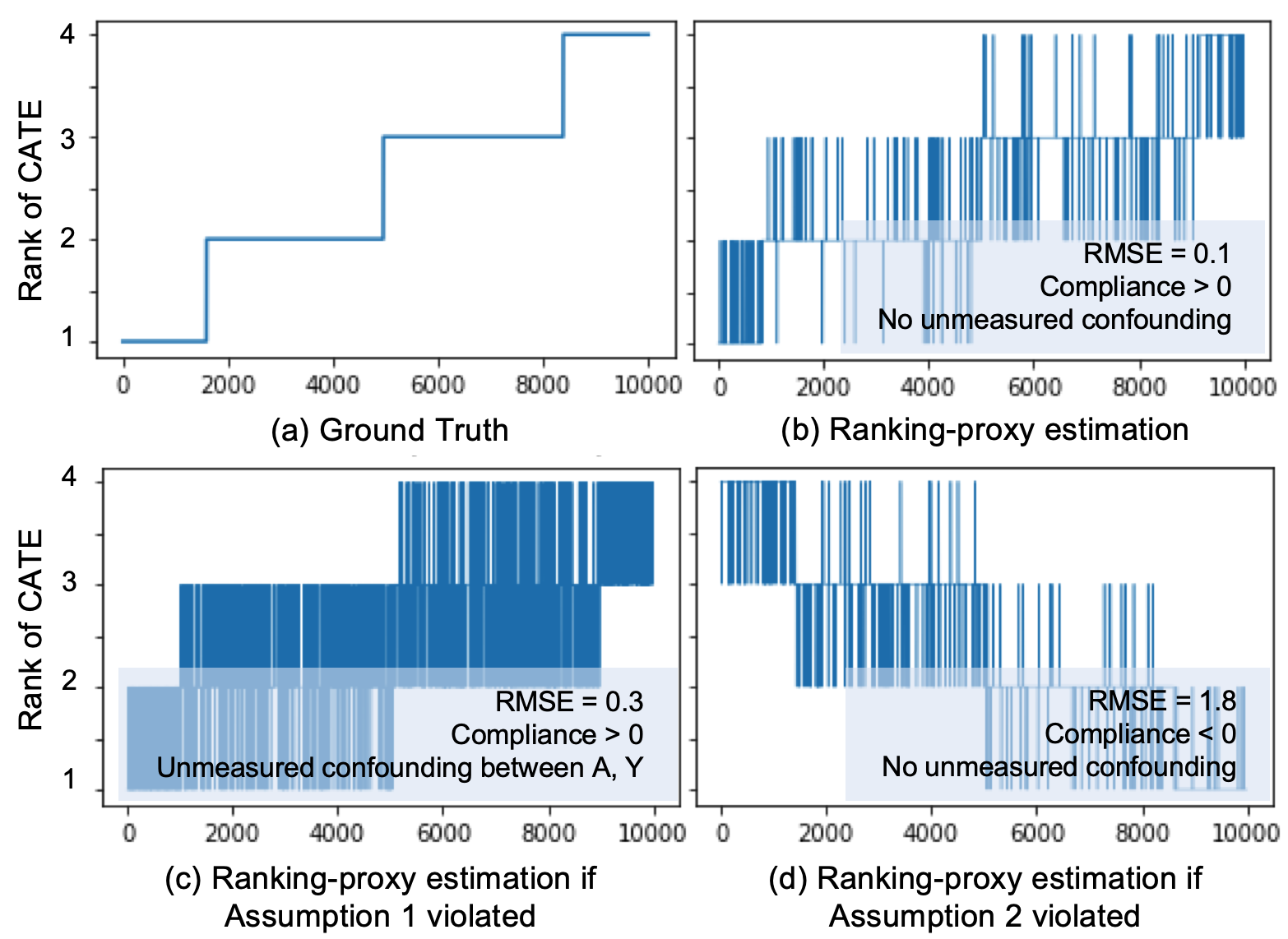}
    \caption{Comparison between the ground-truth rank and the proxy-estimated rank in simulations with or without violations of the two assumptions made in \mname. 
    }
    \label{fig:sim1}
\end{figure}

\section{Simulation}
We conduct a simulation study to demonstrate how \texttt{Split}-\texttt{Treatment} can be used to rank the effect of a target treatment that does not exist in the data. We simulate $10,000$ individuals, characterized by a 50-dimensional variable $\bm{X}$, from a multivariate normal distribution with mean zero. We randomly assign the individuals into 4 groups, and assume that the ground truth CATE of the binary target treatment $Z$ on the outcome $Y$ is $10, 20, 30$ and $40$ respectively in each group. Given each individual $\bm{x}$, the outcome measure $Y^{z=0}$ with no treatment is generated from a linear model over $\bm{x}$ (coefficients are integers randomly picked from 0 to 4 with certain probabilities), and the outcome measure $Y^{z=1}$ with treatment is generated by adding the assigned CATE value to the base measure $Y^{z=0}$. Lastly, we generate a binary proxy treatment $A$ conditional on $Z$ from a contingency table that produces positive compliance for satisfying Assumption 2\footnote{Code for the simulation is provided at \url{https://anonymous.4open.science/r/21143614-e4a1-440e-bbe3-94cbb7d186e4/}}.

The step function in Figure \ref{fig:sim1}(a) shows the ground-truth ranking of CATE of the target treatment $Z$. After masking $Z$ from the simulated data, we fit a doubly robust model with outcome model being a linear regression (namely IPTW-LR), and obtain estimates of ITE for the proxy treatment $A$. We rank the ITE estimates and group them into $4$ equally sized buckets assigned to level 1 to 4. We compute the root mean squared error (RMSE) between the estimated rank and the ground truth rank. Figure \ref{fig:sim1}(b) show the estimated ranks; the obtained RMSE for our IPTW-LR method is low ($0.1$). 

Next we study how the estimation would be biased if we violated the two untestable assumptions made in Section \ref{sec:formulation}. 
To create a dataset that violates Assumption 1, we introduce a confounding variable $U \sim \mathcal{N}(0,1)$, generate outcome measure $Y^{z=0}$ based on $X$ concatenating $U$, and then mask $U$ from the data. We fit the same model above, plot the estimated proxy rank in Figure \ref{fig:sim1}(c) showing that it obtains a larger RMSE of $0.3$. Similarly, for violating Assumption 2, we manipulate the contingency table of $A$ given $Z$ to produce negative compliance and fit the same model to the new simulated data. We plot the estimated proxy rank in Figure \ref{fig:sim1}(d) 
showing that rank was reversed and RMSE increases to $1.8$. This analysis shows the importance of the identifying assumptions (especially Assumption 2) for obtaining the correct estimate.

\begin{figure}[tb]
    \centering
    \includegraphics[width= 0.8\columnwidth]{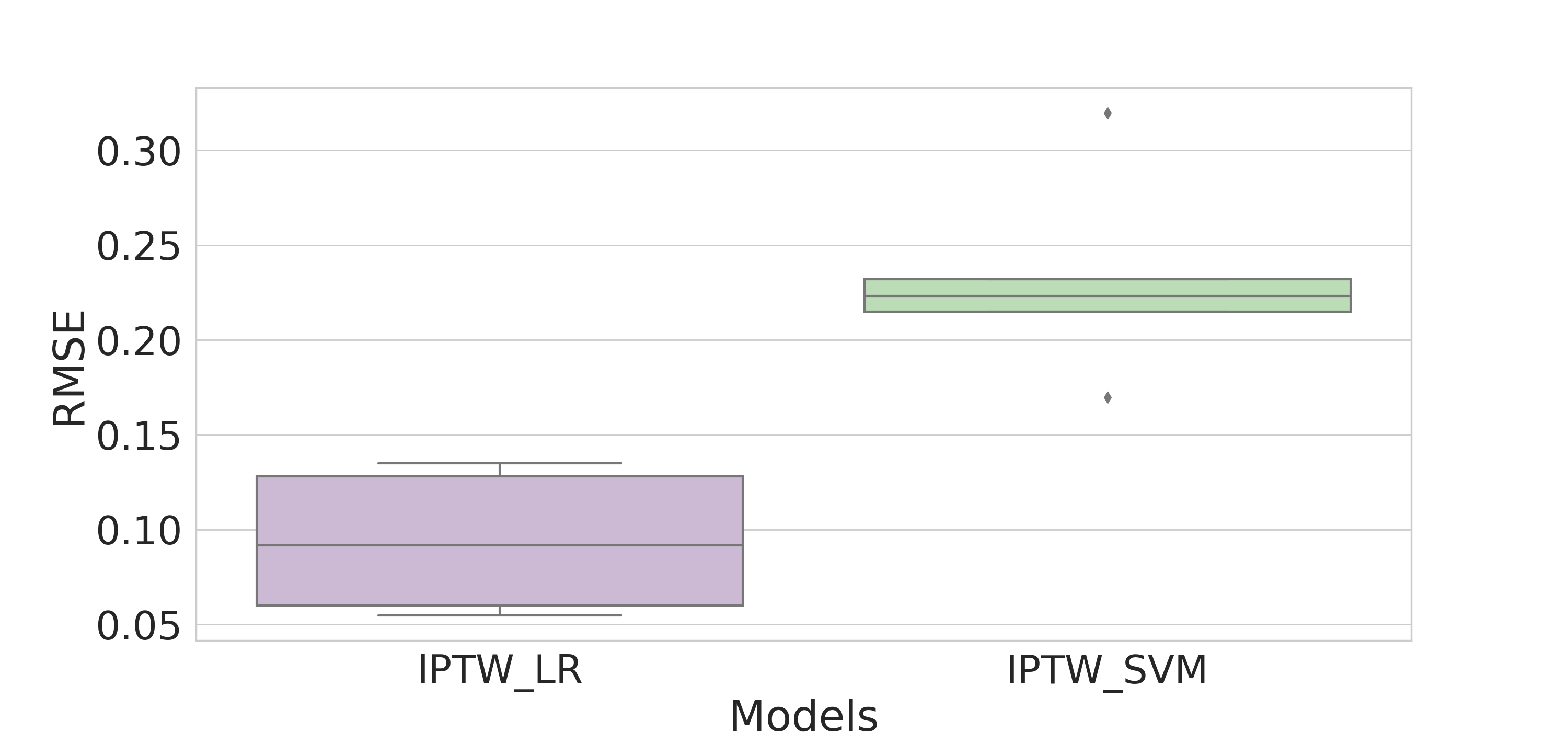}
    \caption{\textit{Unobserved-confounding} analysis. Comparing RMSE between estimated causal effect with and without unobserved confounding, for two causal models. IPTW-LR is less sensitive to unobserved confounding. Box plots are for 5 runs with different degrees of confounding.}
    \label{fig:sim2}
\end{figure}

Finally, we demonstrate the value of  sensitivity analysis for model selection. Consider another doubly robust method that uses Support Vector Machine (SVM) as the outcome regression (namely IPTW-SVM). We use the placebo and the unobserved confounder methods described in Section~\ref{sec:refutation} to test sensitivities of the two methods. We find that placebo test does not help us differentiate between the two models, as both of them give high RMSEs ($1.6$ vs. $1.4$ for IPTW-LR and IPTW-SVM respectively) which is expected due to the placebo treatment. We plot results from the unobserved confounder method in Figure~\ref{fig:sim2}, in which we vary the hyper parameters defined in Eq. (\ref{eq:ref-analysis}) for changing the amount of effect from $U$ to   $A$ and $Y$. Since ground-truth CATE is not available in practice, we compare the two models based on RMSEs between  the {\em estimated} proxy rank under the no-confounder case and those from 5 different simulations under the confounder case. IPTW-LR  obtains lower empirical RMSE than IPTW-SVM and thus it is less sensitive (more robust) to the varying degrees of confounding. This result indicates the IPTW-LR  should be chosen. Sensitivity analysis , therefore allows us to select between competing models for estimating CATE.

\section{Application to a Real data set}
\subsection{Dataset}
We apply our method to feature and product recommendations in a large software ecosystem.  Feature and product recommendations are occasionally shown within the ecosystem as short, clickable messages that lead to web pages that educate individuals about features they already have access to in the ecosystem, and new software products that are available for download.  
We report on results studying a specific product recommendation encouraging individuals to try a new software product.
Prior to running a recommendation campaign, we analyze existing logged user data through the \mname method to identify individuals most likely to benefit from the software product.  In this scenario, the treatment $Z$ is the message that encourages individuals to try the software.  The \mname or proxy treatment $A$ is the individual's first usage of the software during the treatment window.  The outcome $Y$ is sustained usage of the software during the post-treatment window.


%
We run our method on an observational data collected from $2.2M$ users within the ecosystem. For the purpose of validation, we ran a separated randomized experiment on another $1.1M$ users who are randomly exposed to the treatment $Z$, but had not used the recommended software during the pre-campaign window.  The campaign was run from March 29 to April 27 in 2019, during which $66.1\%$ were randomly exposed to $Z$.  Exposure to the treatment $Z$ caused a 7.5\% increase in software adoption. 
We align the timeline between the two datasets, take both pre-treatment and treatment window to be 30 days, and obtain the outcome measure of sustained usage of the recommended software in varying from 0 to 30 days during the post-treatment window.  

\subsection{Feature extraction}
We first extract two sets of snapshots on each day within the pre-treatment window. One snapshot contains $25$ features summarizing the users' total usage in the past 4 weeks of software within the ecosystem and collaboration patterns. The other contains $106$ of similar features but finer grained usage information from software ecosystem. As a result, we reach at two static feature sets, namely \textit{Features}$-25$ and \textit{Features}$-106$, by using only the last day's snapshots in the pre-treatment window; and two dynamic feature sets, namely \textit{Features}$-25-$Seq and \textit{Features}$-106-$Seq, by aggregating (i.e., taking average, min, max, number of increases and decreases) the 30 snapshots within the pre-treatment window. 



\subsection{Baselines and Algorithmic Details}
\label{sec:baselines}
Given each feature set, we explore the following models for outcome regression. Each of these models can be used in a causal mode, addressing confounding issues through use of an IPTW adjustment, or in a predictive mode that does not do such adjustment. Model hyperparameters were tuned via grid search by splitting a $10\%$ validation set from the training data.
\begin{itemize}
    \item Fast-Tree Regression (FTR): An efficient tree regression with gradient boosting \cite{rashmi2015dart}.
    \item Fast-Forest Regression (FFR): An efficient random forest regression using the Fast-Tree learners. 
    \item Poisson Regression (PR): A linear regression with respect to minimizing Poisson loss instead of mean squared errors. 
    \item Convolutional Neural Network (CNN): A 2-layer 1-D convolutional network with Poisson loss.
\end{itemize}

\subsection{Results}
\subsubsection{Evaluation on propensity scores}
We build a logistic regression propensity model that predicts the likelihood that a user gets treated over each of the four feature sets respectively. For each propensity model, we discard the extreme propensity scores they generate that are below the $.01$ quantile or above the $0.99$ quantile.
Comparing the treatment and control group distributions, we find the two groups are well overlapped across all the four feature sets.
%
%
The purpose of propensity score model is for ensuring balance of potential confounds.  We compute the standardized mean differences (SMD) \cite{stuart2010matching} across the features in the treatment and control groups to evaluate the balance. SMD calculates the divergence of the mean feature values between two groups as a fraction of the summed standard deviation of the two groups. 
To validate that our propensity weighting is reducing the confounding, we compare the computed SMDs before IPTW weighting and after.
We find that the SMDs are all reduced after weighting, indicating that IPTW is reducing confounding.  Moreover, we find that, with the exception of one feature, the SMDs of IPTW-weighted Features-106 and Features-106-Seq are all reduced to lower than $0.2$, considered a threshold for well-balanced covariates \cite{stuart2010matching, kiciman2018using}).  



\subsubsection{Evaluation on outcome prediction}
\begin{figure}
    \centering
    \includegraphics[width= .5\textwidth]{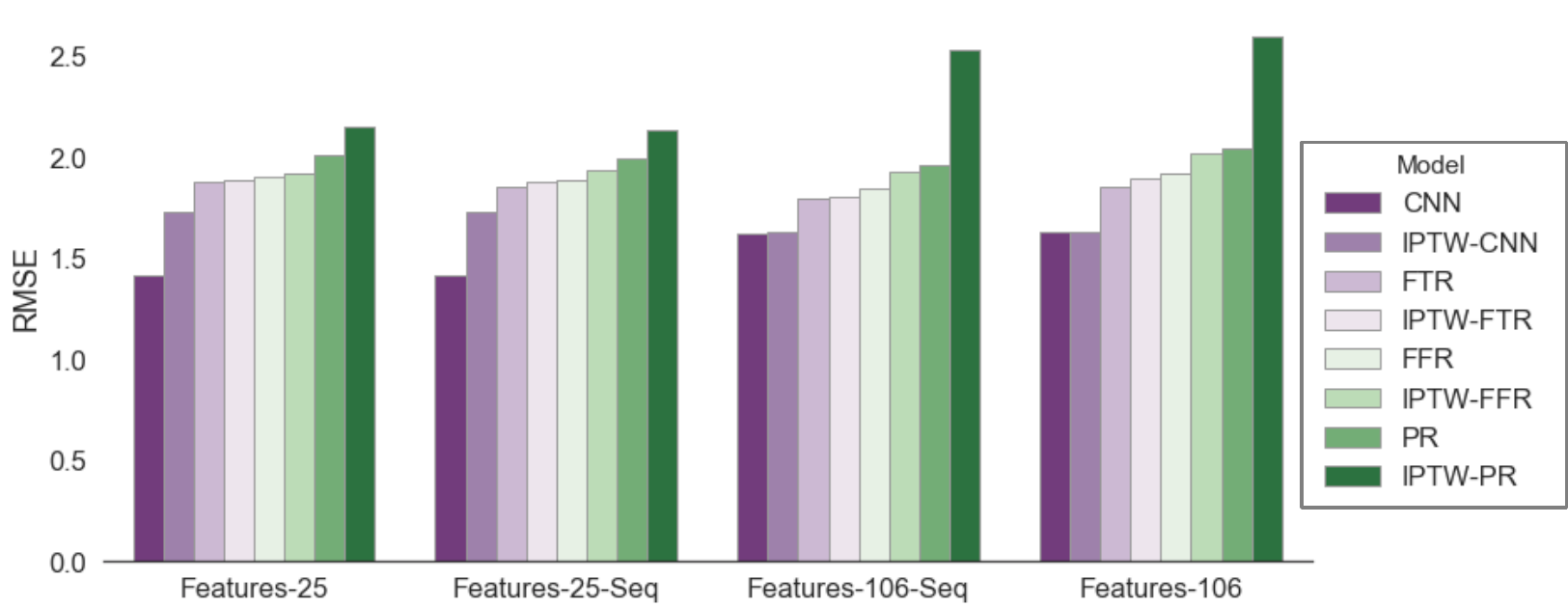}
    \caption{RMSE of outcome predictions on target data from the baseline models over different feature sets; the range of the outcome measure $Y$ varies from 0 to 30.}
    \label{fig:outcome_eval}
\end{figure}

Now we evaluate how accurate the regression models listed in Section \ref{sec:baselines} can generalize their outcome predictions to an independent target data. In Figure \ref{fig:outcome_eval}, we see CNN models give the lowest RMSE across all the four feature sets while Poisson regression models give the highest error; all the other models perform similarly in between. 


\subsubsection{Model Selection via Sensitivity Analysis}
\label{sec:sensitivity}

We first try the placebo-treatment test by replacing the treatment in each dataset with a Bernoulli (p=0.5) random variable. We find that all methods pass the test: no method reports an estimate significantly away from zero for a placebo treatment. For the second test on sensitivity of a method to unobserved confounding, we generate multiple runs over different configurations of the hyperparameters $\alpha$ and $\epsilon$ in Eq. (\ref{eq:ref-analysis}), as below. In choosing these hyper parameters, the goal is to generate a confounder variable $U$ such that it is correlated with both $A$ and $Y$. Different configurations capture different degrees of this correlation, and thus confounding due to $U$.  

\textbf{Features-106}:
\begin{itemize}
    \item $\alpha$: $10^{5}$, $\epsilon$: 40*$\alpha$, Corr. T: 0.38, Corr. Y: 0.18  
    \item $\alpha$: $10^{5}$, $\epsilon$: 100*$\alpha$, Corr. T: 0.25, Corr. Y: 0.12
    \item $\alpha$: $10^{3}$, $\epsilon$: 1700*$\alpha$, Corr. T: 0.53, Corr. Y: 0.25 
\end{itemize}

\textbf{Features-25}:
\begin{itemize}
    \item $\alpha$: $10^{5}$, $\epsilon$: 10*$\alpha$, Corr. T: 0.49, Corr. Y: 0.18
    \item $\alpha$: $10^{5}$, $\epsilon$: 50*$\alpha$, Corr. T: 0.22, Corr. Y: 0.08 
    \item $\alpha$: $10^{3}$, $\epsilon$: 600*$\alpha$, Corr. T: 0.58, Corr. Y: 0.22 
\end{itemize}

\begin{figure}
    \centering
    \includegraphics[width= .5\textwidth]{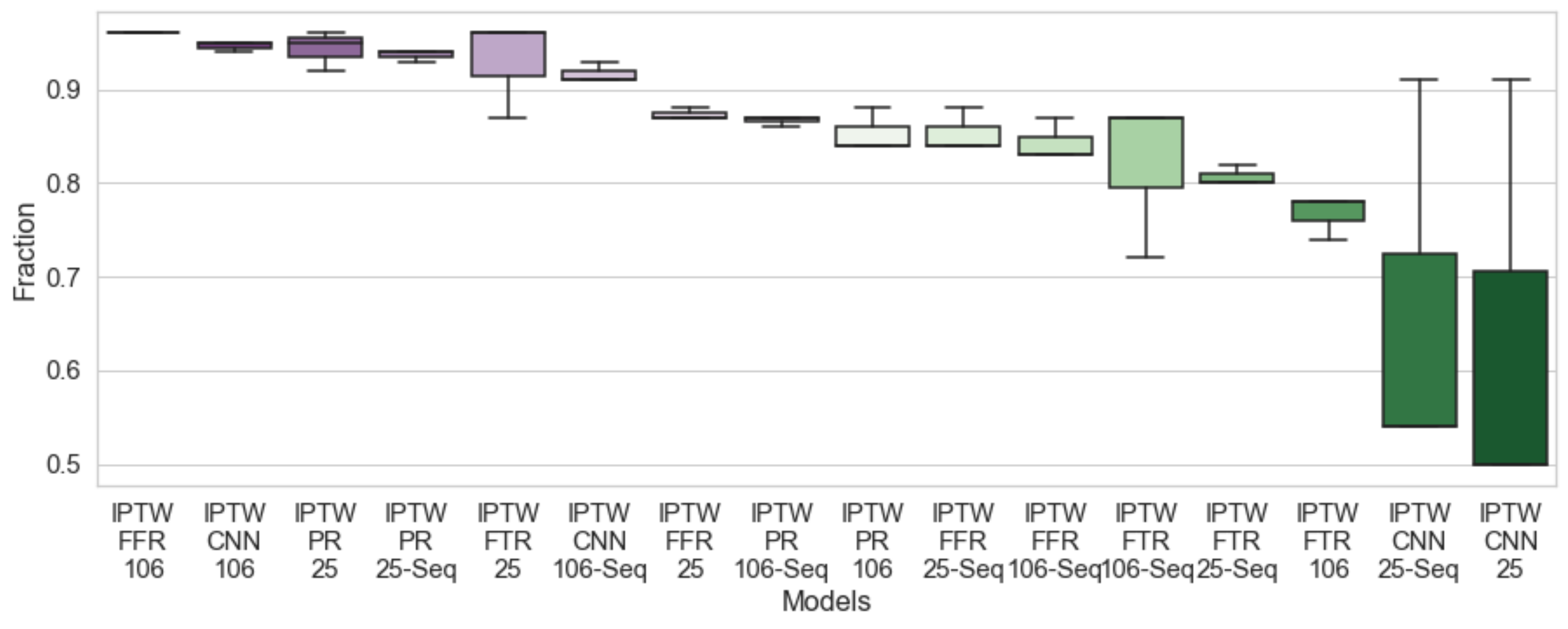}
    \caption{Sensitivity Analysis: Fraction of the top 50-percentile individuals that  remain in the  top 50-percentile after adding an unobserved confounder. Rankings from IPTW-FFR106 and IPTW-CNN106 stay most consistent when unobserved confounding is introduced. Box plots are for 3 runs with different degrees of confounding.}
    \label{fig:sensitivity-analysis}
\end{figure}

We ensure that the correlations across the two feature sets, \texttt{Features-106} and \texttt{Features-25} are similar. Further, we only consider causal models for this analysis since the other methods do not account for confounding adjustment. Figure~\ref{fig:sensitivity-analysis} shows the results across all three configurations, by comparing the fraction of individuals who stay in the top-50 percentile ranking of effect, even after adding the confounder. As reported in Section 4, methods that have a higher fraction of top-50 individuals that are consistent in the original data and the simulated data (with an additional confounder)  are desirable. Note that we do expect that the estimates from each model will change: our goal is to find the methods that change the least with the same amount of additional confounding introduced, and thus are likely to capture more stable relationships. Our first observation is that models based on Features-106 are less sensitive to additional confounding than Features-25, except the IPTW-PR model. The two best models w.r.t. low sensitivity are IPTW-FFR and IPTW-CNN based on Features-106. Among models, we observe models based on IPTW-FFR and IPTW-PR tend to achieve the highest fraction of consistent top-50 individuals, across both feature sets.  While the IPTW-CNN model performs well for Features-106, its sensitivity is the worst among all models for Features-25. 
These results indicate that models based on Features-25 (and models using FTR method) may not be as robust to unobserved confounding as other methods. On balance, we therefore conclude that IPTW-FFR and IPTW-CNN  methods on Features-106 may be most suitable to estimate the causal effect. 

\subsubsection{Validation through Active Experiments}

While we may not frequently have the ability to run active experiments in practice, we do run an A/B experiment in this case, to provide an end-to-end evaluation of the proposed \mname method. 
Through our sensitivity analyses, we have identified that the methods most likely to be reliable are the IPTW-FFR on Features-106 and IPTW-CNN on Features-106; and the worst models are likely to be IPTW-CNN on Features-25-Seq and IPTW-CNN on Features-25.  While we focus the presentation of our results on these best and worst models, our evaluation of the rankings produced by other models is consistent with their quality as identified by our sensitivity experiments. 

%
 The treatment assignment randomized in the A/B test is the message treatment $Z$, while our observational study provides an ITE estimate of the effect of the split-treatment, $P(Y|A)$.  To validate the quality of the rankings implied by our observational models, we treat our randomized experiment as an  instrumental variable (IV) analysis \cite{angrist1996identification}: the message $Z$ acts as a (strong) instrument in our problem because 1) it encourages $A$ the choice to first tryout of the software (with treated individuals being 7.5\% more likely to do $A$ than untreated users);
and 2) its effect on $Y$ the sustained usage of the software is also fully mediated via its effect on the choice of $A$ (known as \textit{endogeneity} ). For further details of IV analysis see~\cite{angrist1996identification,angrist1995two}.  

As an IV experiment does not allow us to calculate ITEs, we use the following strategy to validate the rankings implied by our observational studies.  First, we apply the ITE prediction model learned from our observational study to predict the ITEs for individuals observed in the course of our randomized experiment.  We then split individuals into two sets, $\mathcal{G}_{\text{high},k}$ and a $\mathcal{G}_{\text{low},k}$.  $\mathcal{G}_{\text{high},k}$ consists of all individuals having an ITE in the top $k^{th}$ percentile of predicted ITEs, and $\mathcal{G}_{\text{low},k}$ includes all remaining individuals.  Note that the average treatment effect in $\mathcal{G}_{\text{high},k}$ is necessarily higher than the average treatment effect in $\mathcal{G}_{\text{low},k}$. Secondly, we use our IV experiment to calculate the conditional average treatment effect (CATE) for the both $\mathcal{G}_{\text{high},k}$ and $\mathcal{G}_{\text{low},k}$.\footnote{The CATE, when conditioned on group membership, may also be known as a Local-ATE over the group.}  By conditioning on instrument $Z$ being randomized, we can fit a two-stage least squares (2SLS)~\cite{angrist1995two} to obtain the CATE estimates $\text{CATE}^{(a)}_{\mathcal{G}_i}$ within a specified group $\mathcal{G}_i$.

%
If the ranked causal effects given by our observational study model are accurate, we will expect, for any splitting threshold $k$, that $\text{CATE}^{(a)}_{\mathcal{G}_{\text{high},k}}$ will be higher than $\text{CATE}^{(a)}_{\mathcal{G}_{\text{low},k}}$.  Thus, despite the challenges of gaining ground-truth information about treatment effects, this procedure effectively allows a validation of our model's estimated rankings of the causal effect of the split-treatment $A$.

\begin{figure}[!ht]
 \subfloat[Best model: Our validation of IPTW-FFR on Features-106 (left) shows that CATE for {\em low} and {\em high} groups are correctly separated across all values of $k$.  The non-causal counterpart, FFR on Features-106, is not well-separated. \label{fig:best}]{%
   \includegraphics[width=0.5\textwidth]{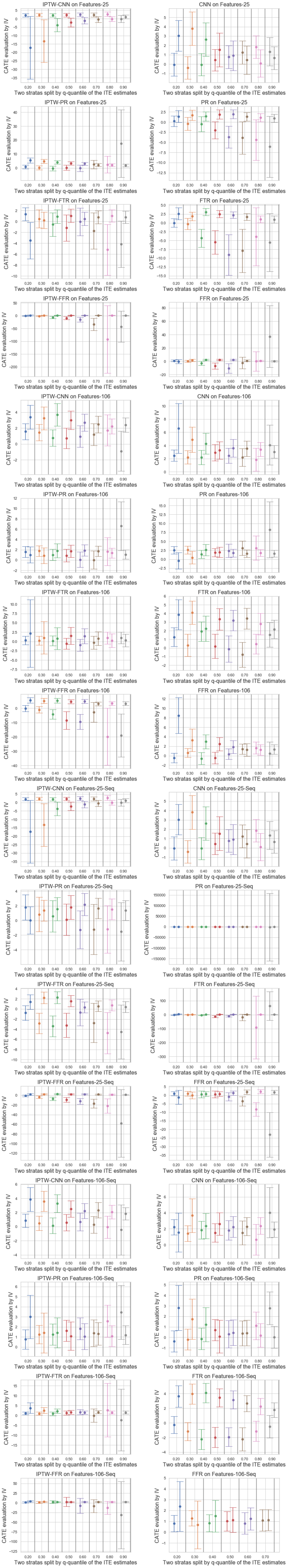}
 }
 \hfill
 \subfloat[Worst model: Our validation of IPTW-CNN on Features-25-seq (left) shows that CATE for {\em low} and {\em high} groups are not correctly separated most values of $k$.  The non-causal counterpart, CNN on Features-25-seq, is worse. \label{fig:worst}]{%
   \includegraphics[width=0.5\textwidth]{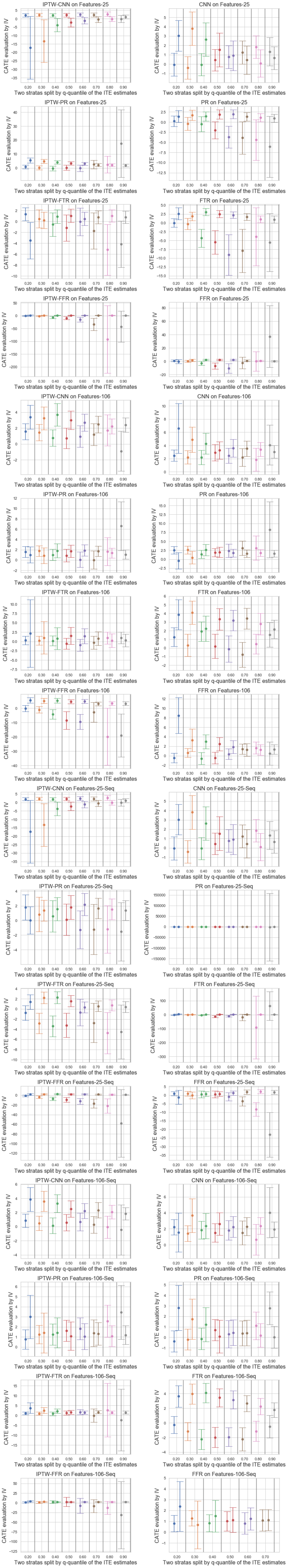}
 }
 \caption{Our CATE evaluation of the best model vs. the worst model chosen by sensitivity analysis (left), as well as their non-causal counterparts (right).  The x-axis represents the threshold $k$, and the y-axis the CATE estimate.  For each $k$, we show the CATE estimate, with standard error, for the paired {\em low} and {\em high} groups in the same color.}
  \label{fig:cate_res}
\end{figure}

Figure~\ref{fig:cate_res} (a) shows the results of this validation procedure over our best causal model, as well as its non-causal counterparts.  We see that the best model consistently ranks the {\em low} and {\em high} groups correctly.  That is, for each threshold $k$, $\text{CATE}^{(a)}_{\mathcal{G}_{\text{high},k}} > \text{CATE}^{(a)}_{\mathcal{G}_{\text{low},k}}$.  We see that this ranking holds particularly strongly at high thresholds.  The quality of our ranking at high thresholds is particularly important in our recommendation domain as, given the multitude of possible feature recommendations, we are likely to display any particular message only to a small portion of users who are most likely to benefit.  
Figure~\ref{fig:cate_res} (b) shows the results of our validation over our worst model, identified as such by our sensitivity analyses.  We see that the worst model clearly fail our experimental validation procedure. Note that the non-causal counterparts for each of these models perform worse in each case.
Validation results for our other observational models (not shown) is consistent with the above results for our best and worst models---i.e., not as good as our best models, but not as bad as our worst models.  
%
Overall, we find that our validation procedure confirms that our refutation and sensitivity analyses are useful in helping to identify the most likely best models based on observational data.

\section{Conclusion}

We presented a practical, observational analysis method for identifying individuals likely to benefit from a novel message or recommendation $Z$ that encourages people to take action $A$. Through causal analysis of existing logs that contain observations of $A$, though not $Z$, we identify people who benefit from $A$, as measured by a target outcome $Y$.  Under a simple assumption that $Z$ is a positive encouragement that increases the likelihood of $A$, this allows targeting of the message $Z$ to individuals most likely to benefit.
A key contribution of our analysis is that our use of refutation tests and sensitivity analyses enables a principled {\em a priori} identification of the feature selection and algorithmic design most likely to be robust and accurate.  We validate our analysis procedure with an A/B experiment in a large real-world setting. 

Promising future work includes development of additional refutation and sensitivity analyses to provide further protection against validity threats;  using characteristics of the identified individuals to aid writers and marketers  in the crafting of messages that better express the benefits individuals may receive; and expansion of our procedures to jointly analyze multiple treatments and outcomes.






\bibliographystyle{ACM-Reference-Format}
\bibliography{reference}

\end{document}